\documentclass[
reprint, longbibliography,
amsmath,amssymb,
 aps,%nobalancelastpage
]{revtex4-1}

\usepackage{braket}
\usepackage{graphicx}% Include figure files
\usepackage{dcolumn}% Align table columns on decimal point
\usepackage{bm}% bold math
\usepackage{amsmath}
\usepackage{float}
\usepackage{comment}% for commenting/hiding text
\newcommand{\norm}[1]{\left\lVert#1\right\rVert}

\begin{document}

%\preprint{APS/123-QED}

\title[Submitted Manuscript]{Radiative decay of an emitter due to non-Markovian interactions with dissipating matter}% Force line breaks with \\

%\begin{comment}

\author{Kritika Jain}
\author{Murugesan Venkatapathi}%
\email{murugesh@iisc.ac.in}
\affiliation{Computational and Statistical Physics Laboratory, Indian Institute of Science, Bangalore, 560012}

%\end{comment}

%\date{} % It is always \today, today,

\begin{abstract}
It is known that the more tractable Markovian models of coupling suited for weak interactions may overestimate the Rabi frequency notably when applied to the strong-coupling regime. Here, a more significant consequence of the non-Markovian interaction between a photon emitter and dissipating matter such as resonant plasmonic nanoparticles is described. A large increase of radiative decay and a diminished non-radiative loss is shown, which unravels the origin of unexpected large enhancements of surface-enhanced-Raman-spectroscopy (SERS), as well as the anomalous enhancements of emission due to extremely small fully absorbing metal nanoparticles less than 10 nm in dimensions. We construct the mixture of pure states of the coupled emitter-nanoparticle system, unlike conventional methods that rely on the orthogonal modes of the nanoparticle alone.
\end{abstract}

\maketitle

%\tableofcontents

In the absence of an enclosing cavity, in the weak-coupling regime of a photon emitter and vacuum, Rabi oscillations between the emitter and a proximal resonant object indicates a strong coupling between them \cite{bellessa2004matter, zengin2015matter, Schlather2013splitting}. The increase in decay rates due to the strongly coupled proximal object, is evaluated using the number of additional optical modes available for the spontaneous emission \cite{Vats2002LDOS}. Regardless of the quantization of the emitter, the object or the fields, so far the theoretical partition of the additional optical states into the radiative and non-radiative parts has reflected the classical scattering and absorption efficiencies of this object \cite{chang2007strong, trugler2008strong, van2012spontaneous, delga2014quantum, pelton2019strong}.

\begin{figure}
		\centering
		%\hspace*{-2mm}
		\includegraphics [width=0.4\textwidth]{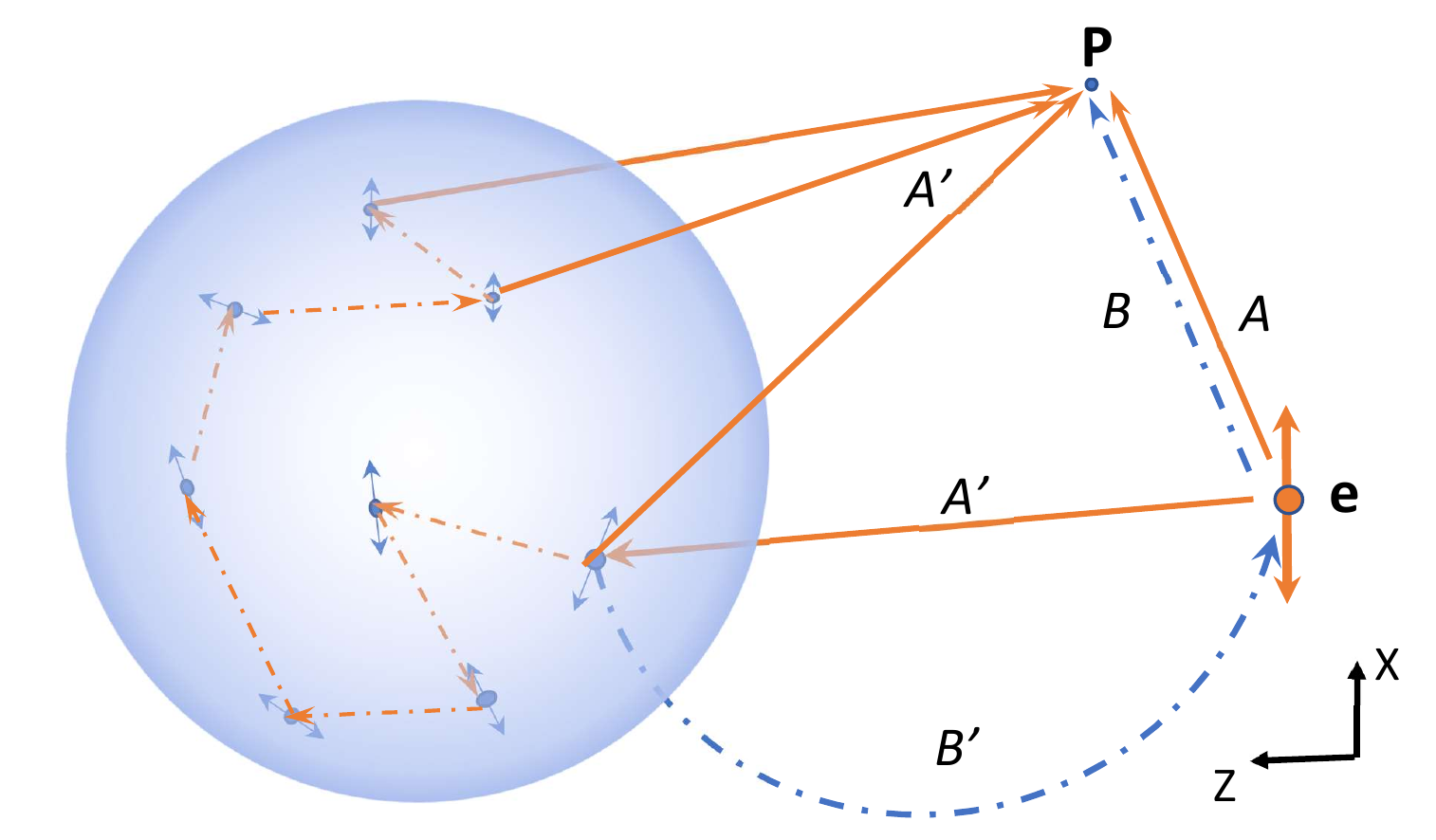}
		\caption{Coupling of emitter \textbf{\textit{e}} with the polarizable matter of a spherical particle and resulting coherent paths of the photon to a far-field point \textbf{\textit{P}}. Superposition of paths $A$ and $A'$ from the emitter and the nanoparticle respectively, represents the Markovian approximation. For stronger couplings, $B'$ (in blue) may return a photon to the emitter and the superposition of multiple excitations renders the process non-Markovian.}\label{fig:Figure1}
\end{figure}

We can recast the coupling of an emitter and a nanoparticle as a quantum interference due to the many paths of a photon from the emitter to a point in vacuum \cite{macovei2007strong,shatokhin2005coherent,safari2019plasmon-phase, Evangelou2009interference}. The paths of a photon from the emitter in the presence of the nanoparticle are described in Figure \ref{fig:Figure1}, where two possible paths of decay are marked as $A$ and $A'$. These two processes represent the memory-less exponentially decaying probability of emission into vacuum, one directly from the emitter, and one from the nanoparticle excited by the photon from emitter. The sum of amplitudes over the multiple paths of the photon to a point $P$, both through the particle and directly from the emitter, represents the weak-coupling (Markovian) approximation. The multiple paths within the particle, including closed loops, result from probable re-absorption and re-emission by constituents of the particle. Numerous experiments have confirmed the weak-coupling predictions of large gains in the radiative decay of the emitter compared to the non-radiative losses, when the larger strongly scattering plasmonic particles ($\gtrsim$ 50 nm in dimension) are placed at optimal distances from the emitter \cite{bardhan2009LDOS, kim2013precise, park2018LDOS}. In typical evaluations, this sum of amplitudes over all the paths $A$ and $A'$ is substituted by a more convenient superposition of the electric fields due to the emitter and and each orthogonal mode of the particle, and summed over these modes. In a related context, the optical theorem for a point emitter also establishes that the additional rate of energy flow due to the above superposition of scattered field of the particle and the direct field from emitter, is equal to the rate of work done on the emitter by the field scattered back from the particle \cite{venkatapathi2012opticaltheorem}. These semi-classical approaches used for determining the increase in local density of optical states (LDOS) due to a weak coupling i.e. the Purcell effect, can be extended to elucidate the case of stronger coupling strengths where the ratio of the Rabi frequency and decay rate is $\gtrsim$ 1. Figure \ref{fig:Figure1} also shows an additional loop $B'$ that may return the photon from the particle to the emitter, which renders the paths $A$ and $A'$ inadequate for the required superposition. The probable multiple excitations of the emitter and the particle due to these paths $B'$, interfere to result in exponentially damped \textit{oscillatory} decays that characterize the non-Markovian process for stronger couplings \cite{Lodahl2011Markov, Kennes2013NM_dynamics, agarwal2013quantum}. The effect of such interactions of an emitter with a metal surface, on the total decay and the effective Rabi frequencies, have been elucidated earlier \cite{Gonzalez2010Markov, yang2017dissipation}. A model for the oscillatory dynamics of excitation energy transfer was proposed as well \cite{Ishizaki2009non-markov}, but the non-Markovian behavior is more generally addressed in the dynamics of open quantum systems \cite{Breuer2007, wolf2008non-markov, Rivas2010non_markov, Breuer2016non-markov}.

It is shown here that the non-Markovian interaction enhances the radiative decay by large factors, along with a diminished non-radiative decay in the dissipating particle. Earlier, a one loop (first order) correction to the conventional decomposition into radiative and non-radiative parts, was proposed using a phenomenological extension, to account for the non-Markovian effects \cite{jain2019strong}. This correction aimed at moderate coupling strengths is shown here to diverge for larger coupling strengths at smaller separations. The non-Markovian process results in counter-intuitive predictions, for the case of proximal nanostructures that are either strongly absorbing or do not scatter light at all. It includes the unexpected giant enhancements of emission in surface-enhanced-Raman-spectroscopy (SERS) where a strongly absorbing rough metal structure increases the radiation exciting a molecule in the near-field, by a few orders of magnitude, but surprisingly with no apparent absorption of the photons emitted by the excited molecule. This divergence of SERS from first-principle theoretical predictions has been widening for decades, as the reported SERS enhancements grew from $10^4$ to $10^{14}$ \cite{schatz2006electromagnetic, Moskovits2013review, kneipp2016SERSrigorous, Heeg2020SERS}. Meanwhile, anomalous enhancements of spontaneous emission near fully absorbing metal nanoparticles less than 10 nm in dimensions, that do not scatter light, have also been reported \cite{haridas2010controlled,haridas2010photoluminescence,kang2011fluorescence,haridas2013photoluminescence, dutta2019large}. Oscillations in their extinction spectra along with small shifts compared to the conventional predictions due to scattering of electrons at boundaries of these small particles have been known, but these marginal dissipative effects on the permittivity do not explain the above anomalous enhancements of radiative decay \cite{Kawabata1966Early_EELS, Fujimoto1968Early_EELS, Ruppin1982Nonlocal, agarwal1983NonlocalSphere, Ginzburg1984, Kreibig1995, Scholl2012NP_EELS}. We note that the low rates of dissipation in these small nanoparticles result in stronger coupling strengths even at large relative separations, and it necessitates the non-Markovian treatment of radiative decay suggested in this work.

The difficulty in the non-Markovian evaluations is the non-locality of the interactions. Continuing from the path based description in the above paragraphs, we use a single two-level emitter and $n-1$ spatially distributed two-level components to represent the proximal nanoparticle. A superposition of paths based on this distributed interacting system is an explicit description of the non-local process, both in time as well as in space. An initial state of the excited system can be described by the $n$ interacting components as
\begin{equation}
  \psi = \{ c_1\ket{\hat{e}_1}+ c_2\ket{\hat{e}_2}+\dots c_n\ket{\hat{e}_n}\}
\end{equation}
where $c_j^*c_j$ represents the unknown probability of excitation of the two-level component $j$. Here $\hat{e}_j$ represents the canonical basis vectors in $n$ dimensions, and $\sum_jc_j^*c_j=1$ for a known initial state. Using these individual two-level components as the basis for representing the collective system more concisely, $\psi = [c_1, c_2,\dots c_n]$, where the emitter is given the index 1. Radiative decay of an initial state is given by the superposition of the radiative decays of the basis states $\ket{\hat{e}_j}$.

The next section on weak coupling of the emitter with the nanoparticle determines a single initial state for this emitter-particle system, analogous to the conventional approaches. The later section on stronger couplings and non-Markovian interactions, presents a method to build a density matrix representing the mixture of initial states of the emitter-particle system with a non-local excitation. A further orthogonalization of the initial states into pure states allows a weighted sum over the superpositions in each pure state, to evaluate radiative decays. %Note that the decay rates of these individual two-level systems are known and given by the material properties of the nanoparticle and the isolated emitter.

\section{Weak coupling with the nanoparticle}\label{sec:Markovian}

In the regime of weak coupling, the initial state of the excited coupled system is given by
\begin{equation}
  \psi = [\psi^e, \psi^p]
\end{equation}
where the initial state of the emitter, $\psi^e$, is known. Under the Markovian approximation, $\psi^p$ can be evaluated using the possible excitation of the particle due to the emitter with an initial state $\psi^e$.

\begin{comment}
Note that the above approximation also implies that the orthogonal eigenstates of the coupled system are simple tensor products of the eigenstates of the isolated emitter and particle. Given a set of orthogonal basis $\{\phi^e\}$, $\{\phi^p\}$ for the emitter and particle, then the corresponding eigenstates of the coupled system are:
\begin{equation}
  \{\phi^s\} = \{\phi^e\} \otimes \{\phi^p\}
\end{equation}
\end{comment}

Without any significant loss of accuracy, the initial state of the excited particle can also be constructed in the form of $n-1$ polarizable point dipoles using a balance of forces. The weak coupling with vacuum allows an evaluation of the interactions using classical fields in the dipole approximation \cite{Feynman1963118, Tignon95Fermi, Debierre2015Fermi}. The absolute values of self-energy of these representative oscillators provides the probability of excitation of the mutually interacting two-level components. It is implicit that each two-level component has a transition energy of $\hbar \omega_0$. Given the polarization of the dipole emitter, we balance the forces at each dipole using
\begin{equation} \label{eq:coupled-system-weak}
 -\frac{1}{\alpha_j}\mathbf{P}_j + \sum_{k=2, k \ne j}^n\mathbf{G(r_j, r_k)}\mathbf{P}_k = \mathbf{E}_{\text{inc.}} = \mathbf{G(r_j, r_1)}\mathbf{P}_1 
\end{equation}
where the polarizability $\alpha_j$ of a spherical grain in the particle can be determined from its size and the dispersive permittivity of the material, using the Clausius-Mosotti relation \cite{purcell1973DDA} and its extensions to include the lattice dispersion \cite{draine1994discrete}. The maximum size of the grain is determined by the wavelength of emission and material properties of the nanoparticle, and the number of such polarizable oscillators $n$ arranged in a hexagonal close packed form can be increased to meet the error tolerance allowed in the solution \cite{draine1994discrete}. $\mathbf{E}_{\text{inc.}}$ is the field incident on the dipole grains due to the emitter. Solving the above coupled system of equations gives us the polarization $\mathbf{P}_j$ for $j=2,3 \dots n$ representing the particle. The required Green dyad $\mathbf{G}$ are evaluated for a given frequency $\omega$, and is defined in the appendix. The solutions can be weighted and integrated with a line-shape around the resonance frequency of the emitter, $\omega_0$, if required.

After solving the above, the required Markovian superposition of paths $A$ and $A'$ in Figure 1 is obtained using the sum of electric fields at points $\mathbf{r}$ due to all oscillators $j$ including the emitter, and the total decay rate due to the particle is evaluated by the imaginary part of the self-interaction of the emitting dipole.
\begin{equation} \label{eq:radiative-decay}
 \Gamma^r = \frac{\epsilon_0}{2\hbar}\oint \mathbf{E}(\mathbf{r})^2 d\mathbf{r} = \frac{\epsilon_0}{2\hbar}\oint \{\sum_{j=1}^n\mathbf{G(r, r_j)}\mathbf{P}_j\}^2 d\mathbf{r} 
\end{equation}
\begin{align} \label{eq:total-decay}
 \Gamma^{\text{total}} & = \frac{4\pi q^2\omega}{mc^2} \cdot \text{Im}\{\sum_{j=2}^n \mathbf{P}_1 \mathbf{G(r_1, r_j)}\mathbf{P}_j\} + \Gamma_0\\
 & = \frac{4\pi q^2\omega}{mc^2} \cdot \text{Im}\{\mathbf{P}_1 [\hat{G}_{1p}\hat{G}_{pp}^{-1}\hat{G}_{p1}]\mathbf{P}_1\} + \Gamma_0
\end{align}
where the values of charge $q$ and mass $m$ represent an electron. The explicit solution of the coupled system of equations \eqref{eq:coupled-system-weak} containing all the $3\times3$ dyads $\mathbf{G}$, can be rearranged concisely in the form of matrices $\hat{G}$ where the subscripts $p$ refer to the interaction of the elements of the particle, and $1$ represents the emitter respectively \cite{venkatapathi2014collective}; see appendix for the full description. This gives us the decay rate in a form where the self-interaction due to the particle explicitly appears as a tensor in the square brackets of the above equation. The non-radiative decay rate $\Gamma^{nr} = \Gamma^{\text{total}}-\Gamma^r$.

Since this solution from equation \eqref{eq:coupled-system-weak} uses the coupling of $n-1$ oscillators only in the nanoparticle, it can be always rewritten as a weighted sum of the $n-1$ orthogonal modes of the particle if required. For particles of regular shapes like spheres, a more compact set of orthogonal modes can also be analytically constructed  using the vector harmonics of the Helmholtz equation. In the weak-coupling regime, the $n$ oscillators and the analytical vector harmonics provide identical results. But an analytical decomposition into orthogonal modes of the emitter-particle $system$ is not tractable for retarded interactions, and the possible quasi-static solutions become inaccurate for stronger couplings. Whereas this description using a large set of $n$ oscillators allows us to numerically construct the eigenstates for the coupled system even with retarded interactions, and this becomes essential in the next section. Higher the refractive index and size of the particle, larger is the number of optical paths (states) in the particle, and so is the number of orthogonal modes or the number of oscillators $n$, required in the summation. A nanoparticle can also produce a large density of paths when a mode is resonant. Note that this weak-coupling approximation does not include the loop that returns the photon to the emitter as shown in Figure \ref{fig:Figure1}. This possible re-absorption of the photon renders the initial state of system as a mixture.

\section{Moderate and Strong couplings with the nanoparticle} \label{sec:non-Markovian}
Proceeding from the previous section on the Markovian approximation for a weak coupling, we introduce two significant refinements to include effects of the non-Markovian interaction on the radiative decay from the system. The former converges to the latter when separations between the emitter and the dissipating nanoparticle increase, and when the size of the particle is sufficiently large. Firstly, we construct a mixture of initial states to represent the coupled system where the photon can also be re-absorbed by the emitter. Secondly, to obtain the superposition of radiative decays over all the oscillators in an initial state and also the ensemble of initial states, we decompose this mixture into a set of orthogonal pure states.

The solution of the coupled classical system using a balance of forces, in equation \eqref{eq:coupled-system-weak}, is used here to further construct a non-local system that shares a photon \cite{svidzinsky2010cooperative, Agarwal2011Wstates}. %In the weak-coupling solution of \eqref{eq:total-decay}, the modified self-energy of the emitter was evaluated using the inverse of the effective polarizability tensor of the emitter due to the nanoparticle as $\frac{1}{\alpha_1}\mathbf{I}-\hat{G}_{1j}\hat{G}_{jj}^{-1}\hat{G}_{j1}$, where polarizability of the free-space emitter $\alpha_1$ represents its self-energy in vacuum and $\mathbf{I}$ is the unit dyad.
Note that solutions $\mathbf{P}_j$ of the $n$ oscillators and  their self energy components represent a single photon \cite{craig1998molecular, Citrin2004chains}. The off-diagonal and diagonal entries of the self energy matrix normalized by Planck's constant are given by \cite{pustovit2010plasmon}:
\begin{align} \label{eq:strong_initial_states}
\Sigma_{jk} & =\frac{\Delta E_{jk}}{\hbar} - i\frac{\Gamma_{jk}}{2} = \frac{4\pi q^2\omega}{mc^2} \cdot\mathbf{P}_j\cdot\mathbf{G(r_j, r_k)}\cdot\mathbf{P}_k\\
\Sigma_{jj} & = \Omega_{jj} - i\frac{4\pi q^2\omega}{mc^2} \cdot \text{Im}\{\frac{\mathbf{P}_j\cdot\mathbf{P}_j}{\alpha_j}\}
\end{align}

$\Omega_{jj}$ determines the strength of coupling of the isolated oscillators with vacuum, and the conclusions presented here are agnostic to it for weak vacuum-coupling; see \cite{Supplementary} for more details. The real parts of the above symmetric matrix provide the rates of exchange of the photon from the dipole $j$ to another dipole $k$. In the rotating wave approximation valid here as $|\Delta E_{jk}| \ll \hbar \omega_0$, the Rabi frequencies are given by the absolute value of shifts in the energy ($\Omega_{jk}$ = $|\Delta E_{jk}|/\hbar$). The imaginary parts $\Gamma_{jk}$ represent the rates of decay of the photon from dipole $j$ due to the other dipole $k$, and the diagonal entries represents the self-interaction of the excited oscillators due to vacuum. The $n$ collective modes (eigenstates) of the excited system provide us a complete set of initial states, $\psi^i$, and corresponding sets of phases and amplitudes of these oscillators. The mixture of these collective eigenstates that are not necessarily orthogonal, determines the Hermitian density matrix $\rho$ of the system.
\begin{align}\label{eq:sigma-matrix}
& \Sigma \ket{\psi^i} = \lambda_i \ket{\psi^i}\\
& \rho = \frac{1}{\sum_{i=1}^n |\lambda_i|}\sum_{i=1}^n  |\lambda_i|\ket{\psi^i}\bra{\psi^i}
\end{align}

\begin{figure}
		\centering
		\hspace*{-2mm}
		\includegraphics [width=0.5\textwidth]{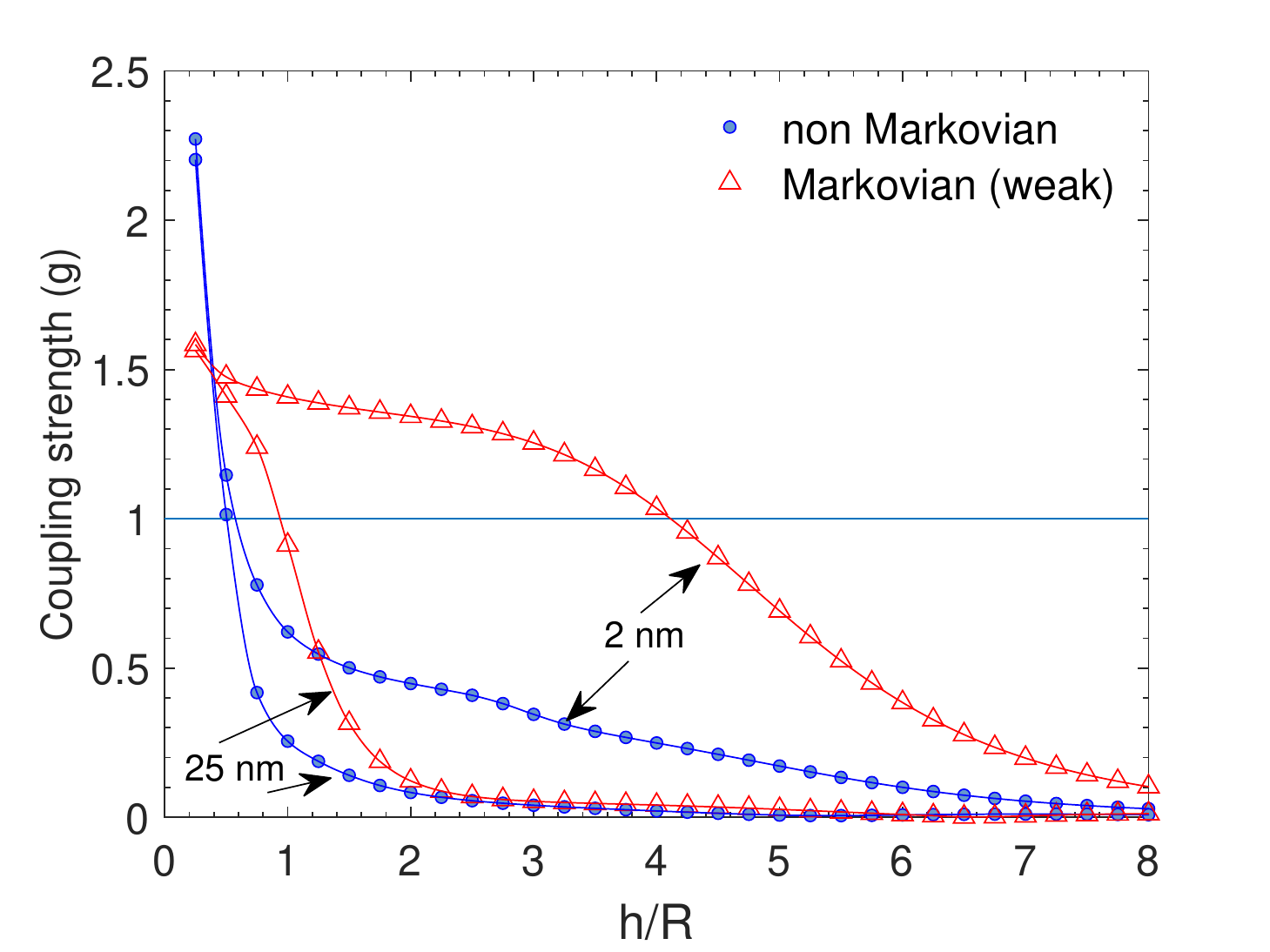}
		\caption{Coupling strength $g$ = $\Omega/\Gamma$ varying with the separation of the emitter $h$ from the surface of gold particles of radii $R$ = 2 nm and $R$ = 25 nm in a medium of refractive index 1.5; $\hbar \omega_0$ = 2.21 eV. Here, $\Gamma=\Gamma^{\text{total}}-\Gamma_0^{nr}$. Initial state of emitter is X or Y polarized. $\Omega$ can be notably different for the two models \cite{Supplementary}.}\label{Figure2}
\end{figure}

The self-interaction of an initial state in the form of its eigenvalue $|\lambda|$ provides its relative weight in the mixture. As the decay rate and Rabi frequency are given by the corresponding imaginary and real parts of $\lambda$, $|\lambda|$ as well represents the relative probability for a configuration of paths given by one of the $n$ initial states. Recall the quantum superposition required over all paths as described in the introduction, which are now represented using a density matrix for the initial states. The ensemble averaged total decay rate and the expected Rabi frequency of oscillation are given by, $\langle\Gamma^{total}\rangle$ = -$2\text{Im}\{\text{tr}(\Sigma)\}$ and $\langle\Omega\rangle$ = $\sum_{i=1}^n |\text{Re}\{\lambda_i\}|$. The predictions of Rabi frequencies can be notably different in the non-Markovian model, as shown by the coupling strengths in Figure \ref{Figure2}.

We decompose the mixture of initial states into a set of orthogonal pure states, and this allows us to sum over the superpositions of the radiative decay from the oscillators in each orthogonal state, to evaluate the expected $\Gamma^r$. Solving Hermitian eigenvalue problems $\rho \ket{\phi^i} = p_i \ket{\phi^i}$, we have probabilities $p_i$ and the pure states $\phi^i$ in the mixture. The amplitude and the relative phase of an oscillator in the pure state $\phi^i$, is used with the normalized polarization set by the solution of the coupled system in equation \eqref{eq:coupled-system-weak}. The polarization of dipoles $j$ for state $\phi^i$ are given by the corresponding entries of the vector:

\begin{equation}
 \mathbf{P}_j^i= \phi_j^i \mathbf{P}_j/\norm{\mathbf{P}_j}
\end{equation}

The radiative decay rate $\Gamma_i^r$ of a pure state $\phi^i$ is evaluated using the polarization given by the above equation and the conventional integral for the superposition of the radiated field from all oscillators $j$ as in equation \eqref{eq:radiative-decay}. The expected radiative decay rate of the system is given by $\langle\Gamma^r\rangle=\sum_{i=1}^n p_i \Gamma_i^r$. The decay rates and Rabi frequencies can also be used to simulate the exponentially damped oscillatory dynamics of the non-Markovian decay, as shown in the appendix. They can result in multi-exponential decays in ensembles \cite{Lodahl2011Markov, agarwal2013quantum}, but our interest here is limited only to expected quantities and the efficiency of emission.

\section{Results and discussion}
\begin{figure}
		\centering
		\hspace*{-2mm}
		\includegraphics [width=0.5\textwidth]{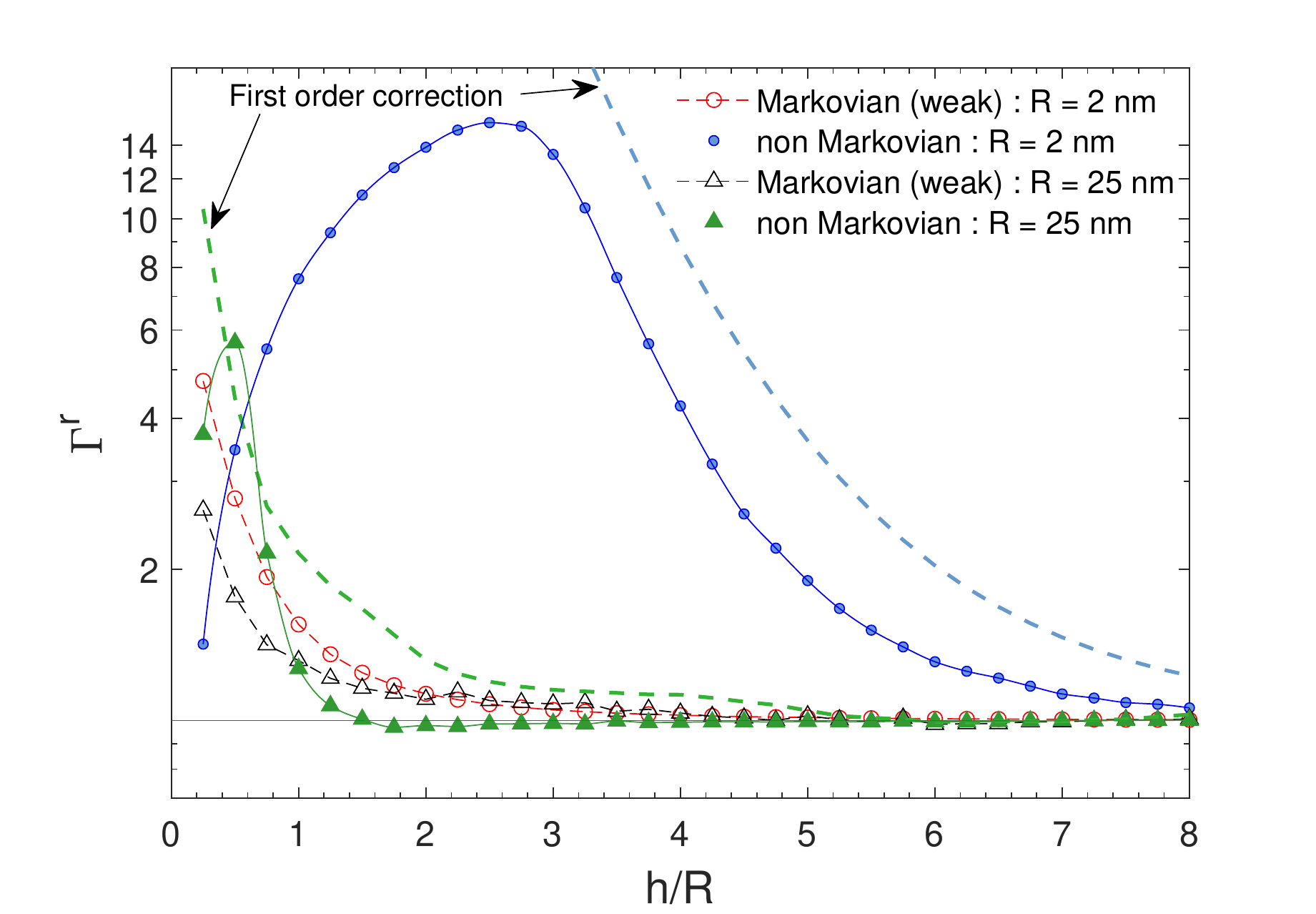}
		\caption{Expected radiative decay rates $\Gamma^r$ normalized by free-space decay rates $\Gamma_0^r$ for emission energy $\hbar \omega_0$ = 2.21 eV at varying separations $h$ from gold nanospheres. Initial state of emitter is X or Y polarized. The first order correction of the Markovian model \cite{jain2019strong} is given by the dashed lines. This correction for effective values was given by $\Gamma_{eff.}^r = \Gamma^r + e^{-1/g} \cdot \Gamma_1^{nr}$, where subscript `1' represents the dipole mode.}\label{Figure3}
\end{figure}

In this section, we compare the results of the Markovian approximation described in section \ref{sec:Markovian}, and the non-Markovian evaluations described in section \ref{sec:non-Markovian}, for gold nanospheres of various sizes and separations from the emitter. The plots of the the radiative decay in Figure \ref{Figure3} show an unconventional large peak for the 2 nm radii fully absorbing metal particles even at large relative separations. The earlier proposed first order correction of the Markovian evaluations suited for weak coupling, captures the large increase of radiative decays, but it diverges for smaller separations and larger coupling strengths. Unlike the Markovian radiative decay rates that seem to diverge very close to the surface of the nanoparticle, the non-Markovian expected decay rates tend to approach a finite value that may be determined using only the permittivity of the surface and $\Gamma_0^r$. This potential consistency of non-Markovian results with the stringent limiting conditions may require further studies for even smaller separations. 

\begin{figure}
		\centering
	%	\hspace*{-2mm}
	%	 \includegraphics[width=1.0\textwidth,trim={2.5cm 2.5cm 1.5cm  1cm},clip=true]{final_QE_main.eps}	
	%	\includegraphics [width=1.0\textwidth]{QE.eps}
        \includegraphics[width=1.0\textwidth,trim={2.5cm 2.5cm 1.5cm  1cm},clip=true]{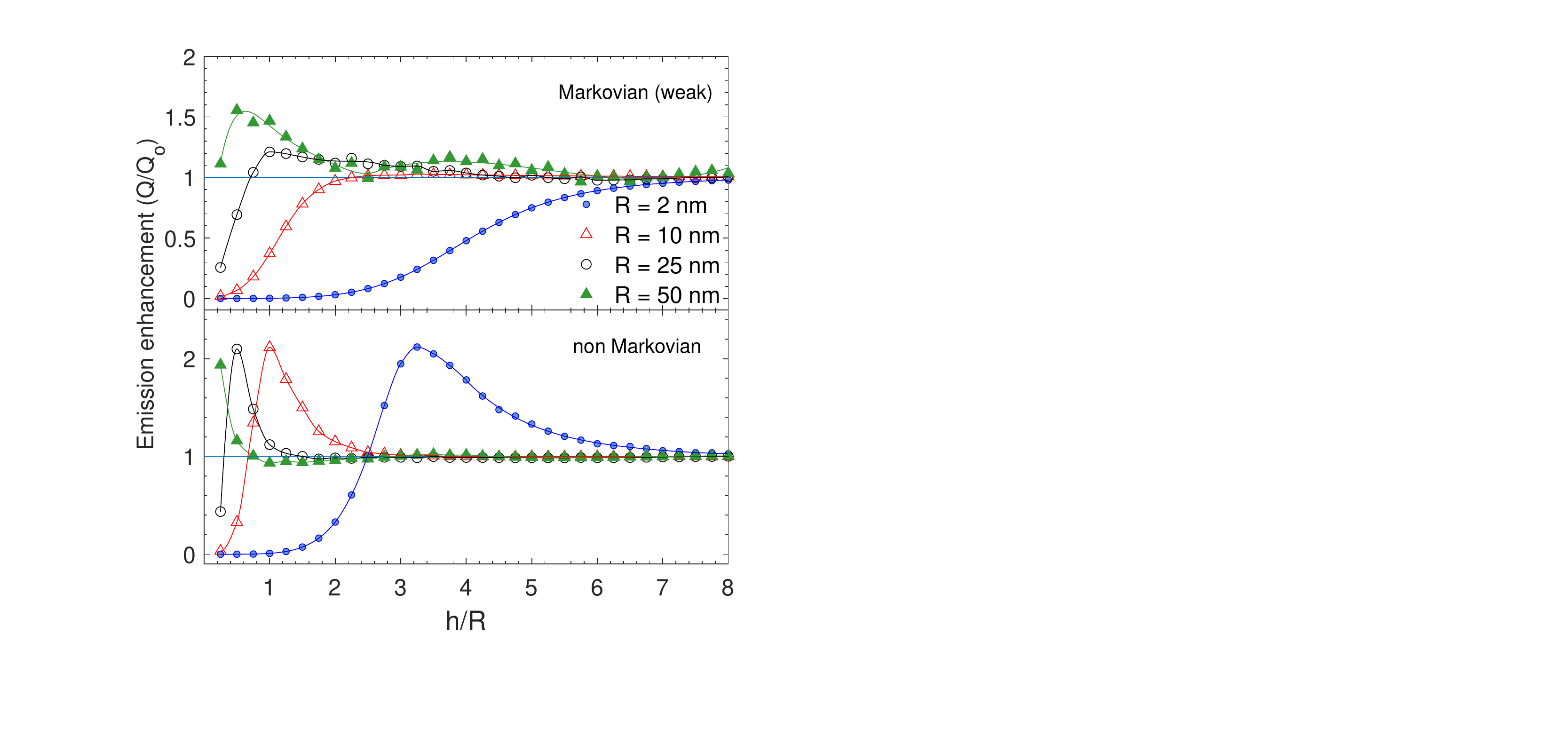}
		\caption{Normalized quantum efficiencies for an X or Y polarized initial state of emitter show enhancement of $Q$=$\Gamma^r/\Gamma^{total}$ for even the smaller 2 nm gold nanoparticles, in the case of the non-Markovian model. $Q_0$=1/3 was assumed; the enhancements have an inverse relationship with $Q_0$.}\label{Figure4}
\end{figure}

\begin{figure}
		\centering
	%	\hspace*{-2mm}
	%	 \includegraphics[width=1.0\textwidth,trim={2.5cm 2.5cm 1.5cm  1cm},clip=true]{final_QE_pol1_S3.eps} 	%	\includegraphics [width=1.0\textwidth]{QE.eps}
    	\includegraphics[width=1.0\textwidth,trim={2.5cm 2.5cm 1.5cm  1cm},clip=true]{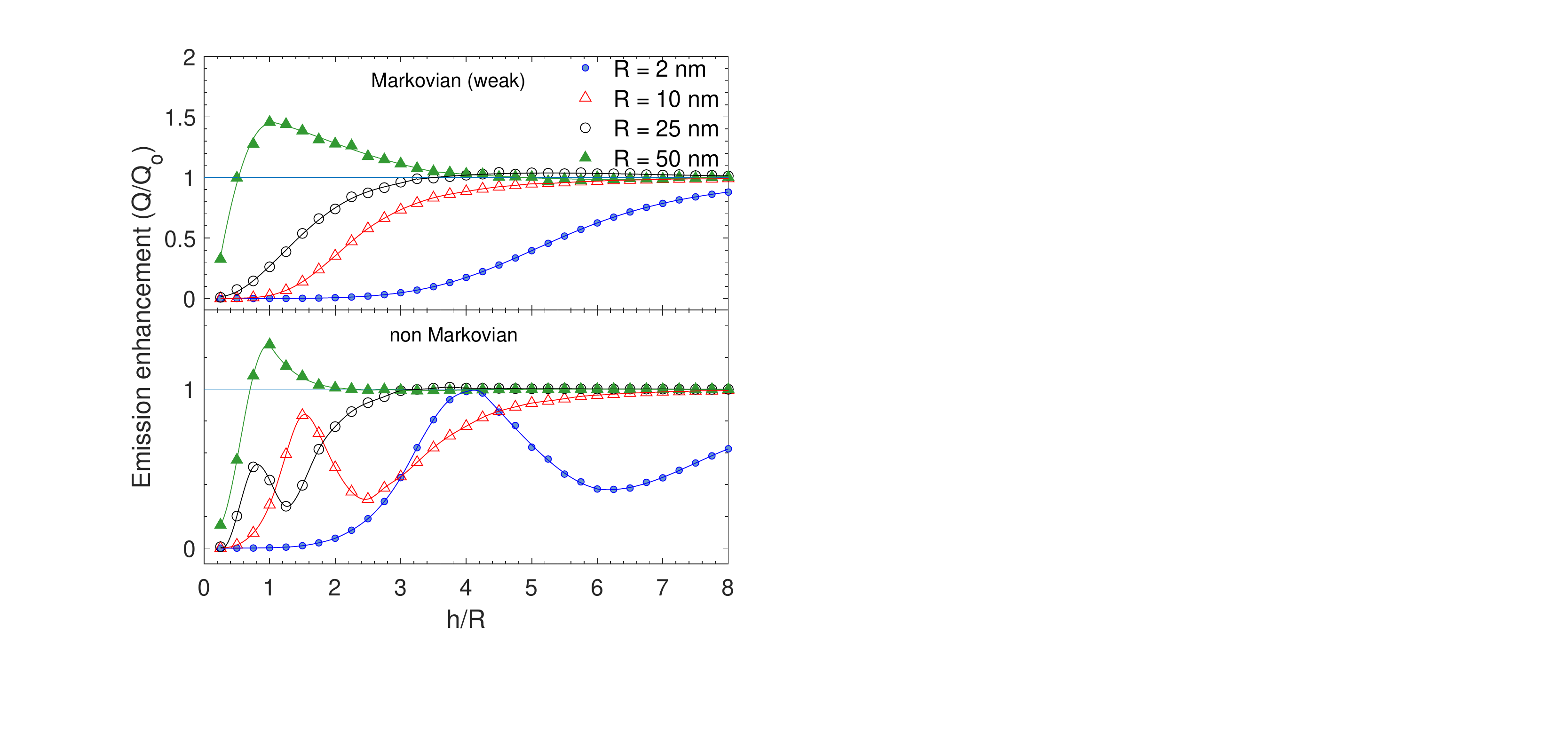}
		\caption{Normalized quantum efficiencies for a Z-polarized initial state of emitter show suppressed quenching for the smaller gold nanoparticle, in the case of the non-Markovian model. $Q_0$=1/3 was assumed, and the enhancements have an inverse relationship with $Q_0$.}\label{Figure5}
\end{figure}

Similarly, the quantum efficiencies shown in Figure \ref{Figure4} for X or Y polarized initial state of emitter contrast the predictions of the Markovian model with its large enhancements for the smaller fully absorbing particles in a medium of index 1.5. The plasmon resonance of the gold nanoparticles in this medium is strong at this energy of 2.21 eV, and the corresponding values for Z-polarization in Figure \ref{Figure5} show suppressed quenching in these cases. The movement of the peak efficiencies towards the smaller separations, and the possible enhancement due to smaller particles, are significant for unraveling the mechanism of SERS. The rough metal surface and its smaller features while enhancing the incident radiation in its near-field, are also shown here to enhance the emission from the molecules at such separations less than 10 nm. The notable quantum efficiency of the non-Markovian evaluations, along with the larger decay rates near the metal surface (i.e. larger ground-state population for the case of a repeated excitation), together can predict the very large gains of Raman signals in the near-field of a metal structure. On the other hand, the conventional models predict a large dissipation (non-radiative decay rates) and a very low quantum efficiency, and this would limit SERS gains to less than $10^5$ even with the near-field enhancement of incident radiation and possible repeated excitations, thus contradicting the experimental observations \cite{jain2019strong}. 

\begin{figure}
		\centering
		\hspace*{-2mm}
		\includegraphics [width=0.5\textwidth]{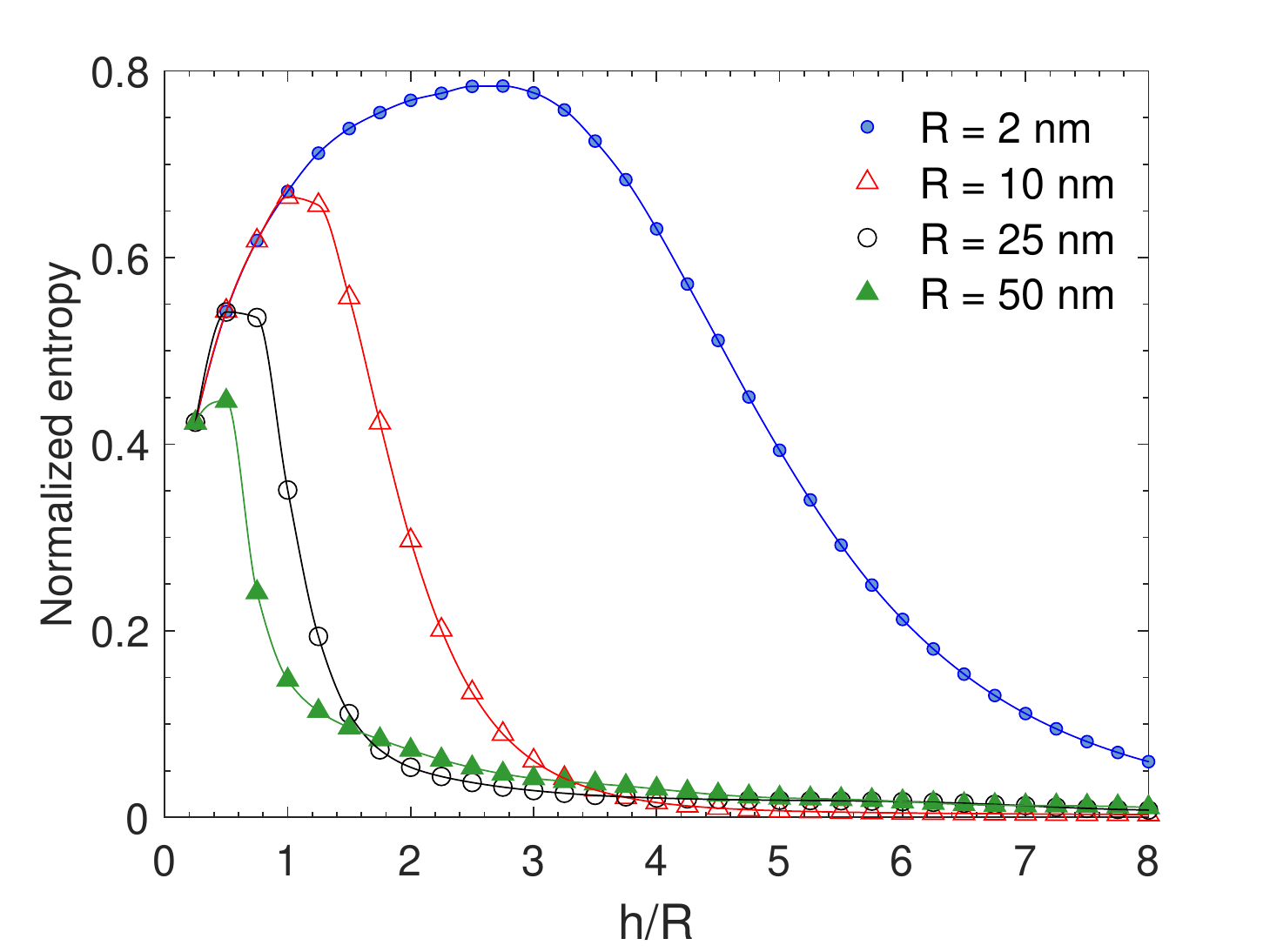}
		\caption{The normalized von Neumann entropy of the mixture, $\frac{-1}{\log n}\sum_1^n p_i\log p_i$, varying with relative separations for gold particles of different radii $R$. Initial state of emitter was X or Y polarized and number of oscillators $n$ is 1419.} \label{Figure6}
\end{figure}

The Markovian models of coupling considered only the uncertainty of the path of the photon from the emitter, and the corresponding interference. In the case of proximal strongly absorbing structures, or fully absorbing (non-scattering) nanostructures, the large effect of the possible re-absorption of photons from the excited nanostructure by the proximal emitter at ground-state, represented by a non-Markovian loop  in Figure \ref{fig:Figure1}, has to be accounted by a mixture of initial states of the system; see Figure \ref{Figure6} for entropy of the mixtures. From a classical perspective, one may relate the origin of this effect to the evanescent fields \cite{adam93evanescent} of the excited dissipating nanostructure, coupling back to the emitter at ground-state.
%\vspace{2mm}
%\begin{comment}

\begin{acknowledgements}
M.V. thanks Girish S. Agarwal for illuminating discussions of the literature on self-interactions. K.J. and M.V. thank the department of Computational \& Data Sciences, Indian Institute of Science for its generous support.
\end{acknowledgements}
%\end{comment}

\appendix
\newcommand\numberthis{\addtocounter{equation}{1}\tag{\theequation}}
\section{Green Dyads}
The required dyads $\mathbf{G}$ for interaction among the point-dipoles, are solutions for a point source in a homogeneous background:
\begin{equation}
\bigtriangledown \times \bigtriangledown \times \mathbf{G}(\mathbf{r,r_j};\omega)- k^2\mathbf{G}(\mathbf{r,r_j};\omega) = \mathbf{I}\delta(\mathbf{r-r_j}).
\end{equation}
giving us
\begin{equation} \label{dipole_green}
\mathbf{G}(\mathbf{r_i,r_j};\omega)=(\mathbf{I}+\frac{\bigtriangledown\bigtriangledown}{k^2})g(\norm{\mathbf{r_i}-\mathbf{r_j}})
\end{equation}
where \(g(r)=\frac{e^{ikr}}{4\pi r}\). $\mathbf{I}$ is a unit dyad and the wave number $k=\sqrt{\epsilon} \frac{\omega}{c}$, and $\delta(\mathbf{r-r_j})$ represents the point source at $\mathbf{r_j}$. The global matrix $\hat{G}$ with indices ranging from $0$ to at most $3n-4$ can be written in terms of the $3\times3$ dyads $\mathbf{G}$ for $j, k = 2,3\dots n$ as
\begin{align*}
&\hat{G}_{pp}(3[j-2]\rightarrow3[j-2]+2,3[k-2]\rightarrow3[k-2]+2)\\
& =\mathbf{G}(\mathbf{r_j,r_k};\omega)\\
&\hat{G}_{1p}(0\rightarrow2,3[j-2]\rightarrow3[j-2]+2)=\mathbf{G}(\mathbf{r_1,r_j};\omega) \numberthis \label{eq:Global_dyads}
\end{align*}
and $\hat{G}_{1p}^T=\hat{G}_{p1}$. Thus, $\hat{G}_{1p}$ is of size $3\times3(n-1)$ and $\hat{G}_{pp}$ is of size $3(n-1)\times3(n-1)$.

\renewcommand{\thefigure}{A\arabic{figure}}
\setcounter{figure}{0}
An example highlighting the distinction between the oscillatory dynamics of the non-Markovian decay profile compared to the memory-less exponential decay, is shown in Figure \ref{fig:dynamics}.

\begin{figure} [H]
		\centering
		%\hspace*{-2mm}
		\includegraphics [width=0.5\textwidth]{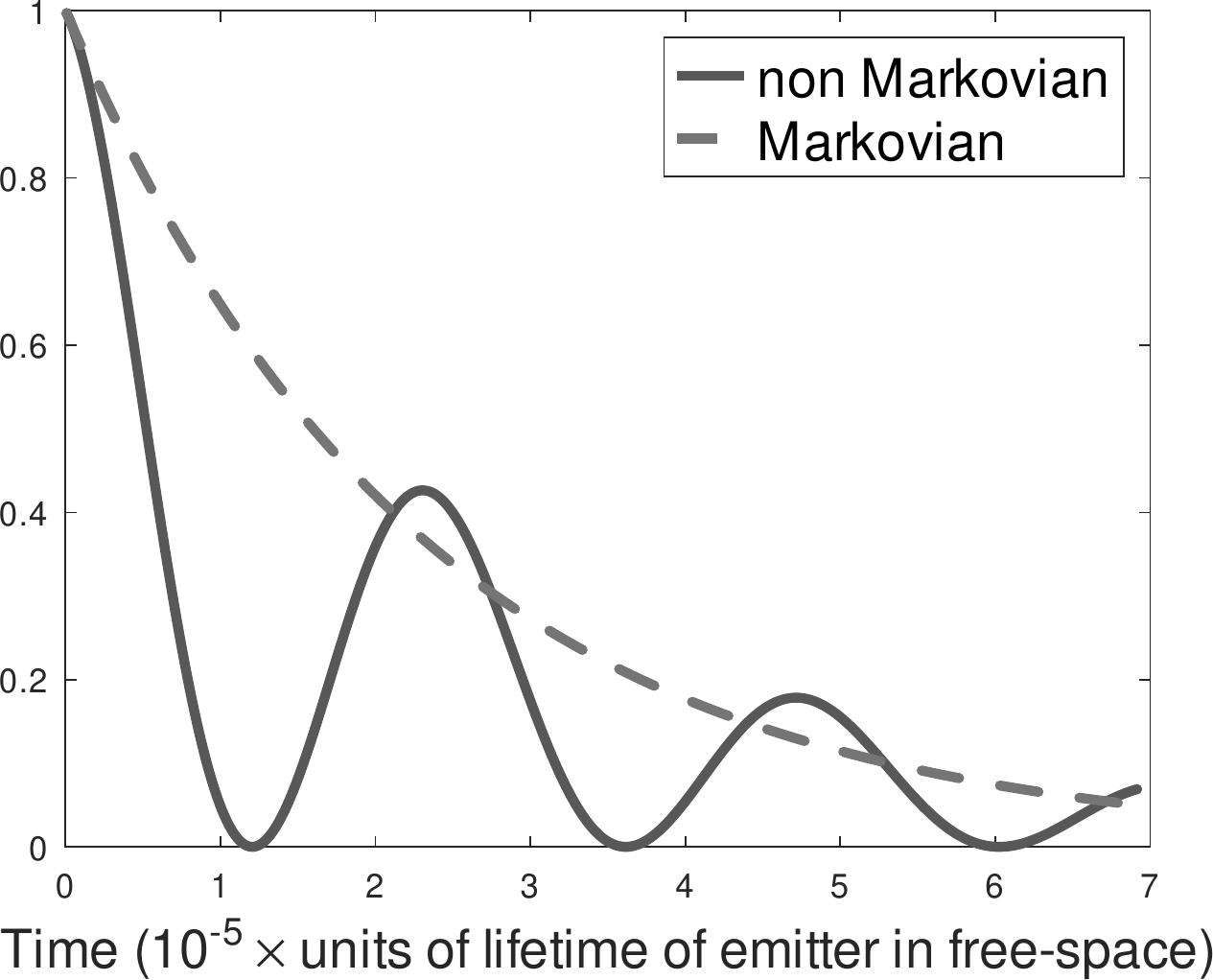}
		\caption{Decay of the coupled system with the separation of the emitter by 1 nm from the surface of a gold particle of radii $R$ = 2 nm. Emitter's initial state is X or Y polarized. Emission energy $\hbar \omega_0$ = 2.21 eV and refractive index of medium is 1.5. Expected total decay rates and Rabi frequencies among the oscillators were used for evaluating the non-Markovian decay profile $\cos^2(\Omega t)e^{-\Gamma t}$ i.e. the probability density normalized with the density at $t=0$. The corresponding exponential Markovian decay profile uses the rate from equation \eqref{eq:total-decay}. An integral of the latter probability density of decay in a limit [0,T] indeed represents an exponentially decreasing probability in $T$ for the excited system, while the integral of the former non-Markovian probability density does not.} \label{fig:dynamics}
\end{figure}

\bibliography{references}

\end{document}